\documentclass{aa}

\bibpunct{(}{)}{;}{a}{}{,} % to follow the A&A style
\usepackage{graphicx}
\usepackage{txfonts}
\usepackage{multirow}
\usepackage{booktabs}
\usepackage{hyperref}
\usepackage{dcolumn}
\usepackage{subfig} 
\newcolumntype{d}[1]{D{.}{.}{#1}} %表示输入为小数点，显示为小数点，参数是根据几位小数来考虑列宽度

\begin{document}

\title{Precision premium transformation -- a high-precision astrometric solution based on the precision premium curve}

\subtitle{}

\author{Z.J.~Zheng
      \inst{1,2},
      Q.Y.~Peng\inst{2,1}\thanks{Contact:tpengqy@jnu.edu.cn}
      \and
      F.R.~Lin\inst{3,1}
      \and
      D.~Li\inst{1}
      \and
      Y.~Zheng\inst{1}
      }

\institute{School of Mathematics and Computer, Guangdong Ocean University, Zhanjiang 524088, China\\
     \and
      Sino-French Joint Laboratory for Astrometry, Dynamics and Space Science, Jinan University, Guangzhou 510632, China\\
     \and
     School of Software, Jiangxi Normal University, Nanchang 330022, China
     }

\date{}

% \abstract{}{}{}{}{}
% 5 {} token are mandatory

\abstract
{In Gaia era, atmospheric
turbulence, which causes stochastic wander of a star image, is a fundamental limitation to the astrometric
accuracy of ground-based optical imaging. However, the positional bias caused by turbulence (called turbulence error here) can be effectively reduced by measuring a target relative to another reference (a star or a fast-moving
target) which locates in the range of only several tens of arcsec, since they suffer from similar turbulence errors. This phenomenon is 
called the precision premium and has been effectively applied to the astrometry of solar system. Further investigation for the precision premium shows that, the precision premium works at less than about 100
arcsec for two specific objects and the relative positional precision as a function of their angular 
seperation can be well fitted by a sigmoidal function, called the precision premium curve (PPC).}
{We want to reduce the turbulence error of a target if it is imaged in an area of high stellar density of a
ground-based observation by taking advantage of more Gaia reference stars. }
{Based on the PPC, 
we proposed a high-precision astrometric solution called precision premium transformation (PPT) in this paper, which
takes advantage of high similarity of turbulence errors in a small region and the dense Gaia reference stars in the region to reduce
the turbulence errors on the observation, through a weighted solution.}
{Through systematic analysis, the PPT method exhibits significant advantages in terms of not only precision but also applicability when a target is imaged in an area of high stellar density. The PPT method is also applied to the determination of the proper motion of an open cluster, and the results demonstrate and quantify benefits that the PPT method bestows on ground-based astrometry.}
{}

\titlerunning{A high-precision astrometric solution based on PPC}
%\authorrunning{<name(s) of author(s)}
\maketitle

\section{Introduction} \label{sec:intro}

The advent of Gaia catalogue~\citep{Gaia2016A&A,Gaia2018A&A,Gaia2021A&A} revolutionizes astrometry and enables many moderate-size telescopes to deliver targets' positions with the precision of tens of milli-arcseconds~(mas) since there are an adequate number of high-precision Gaia reference stars in the field of view~(FOV). To achieve better precision, correction for many other systematic errors needs to be taken into account, and one of the largest errors for ground-based astrometry should result from atmospheric turbulence, which causes stochastic wander of a star image. 
%due to refraction by atmospheric density fluctuations.    

It is found that astrometric error caused by atmospheric turbulence is probably just a random variation, not a significant trend~\citep{Bernstein2017PASP}. Recently, Fortino et al.~(\citeyear{Fortino_2021}) and Léget et al.~(\citeyear{refId0}) propose using Gaussian processes to model the astrometric errors for the whole FOV caused by turbulence, which is a fairly complex method. However, if the target is imaged near another reference~(a star or a fast-moving object) which locates in a range of only several tens of arcsec, it can be measured just relative to the neighbouring reference since they suffer from almost the same positional biases caused by turbulence, called turbulence error in this paper. This phenomenon is called the precision premium, which was first presented by Pascu~(\citeyear{Pascu1994}). To achieve a significantly improved result, it is required that the high signal-to-noise ratio~(SNR) of the target and the reference on the observation. Many high-precision astrometric results of the bright natural satellite observations have been obtained via the precision premium when two bright satellites' separation 
is small enough~\citep{Peng2008P&SS,Peng2012MNRAS}. Based on the idea of the precision premium, a method called mutual approximation is developed. It measures central instant at the closest approach between a fast moving target and another reference in the sky plane and is applied to the astrometry of Galilean satellite~\citep{Morgado2019MNRAS}, Uranus satellites~\citep{Santos2019MNRAS}, and a main-belt asteroid (702) Alauda~\citep{Guo2023MNRAS}. The
best positional precision through the precision premium can reach about 1-3 and 1-4 mas in R.A., and Decl., respectively~\citep{Guo2023MNRAS}.

To further benefit from the precision premium, the effective range of the precision premium is investigated and it is found that the precision premium works at less than about 100 arcsec for two specific objects 
and the relative positional precision can be well fitted by a sigmoidal function~\citep{Lin2019MNRAS}. And it is also found that, the improvement 
of precision due to the precision premium increases almost linearly
with the decrease of the objects' angular separation in a certain range. 

If a target is imaged in an area of high stellar density of a ground-based observation, the astrometric procedure should have the potential to further reduce its suffered turbulence error when taking more Gaia stars as reference stars. For many solar system targets, such case is exceedingly rare. However, astrometry for many distant targets fits the case, such as globular clusters, open clusters, Galactic bulge stars. What is more, their ground-based observations should have been accumulated for many years for some scientific programs and can lead to a more precision answer to a scientific research such as the evolution of the universe if the turbulence errors can be further reduced. 

Based on the precision premium curve (PPC), we proposed a high-precision astrometric solution called precision premium transformation~(PPT) in this paper, which takes advantage of high similarity of the turbulence errors in a small region and the high-precision Gaia reference stars in the region to reduce the turbulence errors on the observations. The PPT method is different with the previous methods in two ways. Firstly, a target would be measured relative to several reference stars, including the high-SNR reference stars and the low-SNR reference stars as well. Secondly, based on the work of Lin et al.~(\citeyear{Lin2019MNRAS}), better astrometric precision would be achieved in a smaller region. However, the determination for the plate model needs an adequate number of reference stars in such a small region. Therefore, the region for measuring a target should be increased gradually until the determination for the plate model is credible, of which the radius are not larger than 100 arcsec. 

Considering the PPT method relies on the density of Gaia stars on the observation, we chose a open cluster, M35, which has been acalibration field of our astrometric programs~\citep{Peng2015MNRAS,Xie2019P&SS,Shang2022AJ} for many years, to test its performance. We would compare the difference between the PPT method and the traditional precision premium (PP) application in terms of not only precision but also applicability. The PPT method is also applied to the determination of the proper motion of an open cluster M35, and the results are compared with the ones using a conventional weighted fourth-order polynomial. 

This paper is organized as follows. In Section 2, the observations and the corresponding instruments 
used to capture them are presented in detail. In Section 3, the procedure of the PPT method is given. In Section 4, the results derived by the PPT method are analysed systematically, through comparing with the traditional PP application and a conventional weighted fourth-order polynomial. In Section 5, as a demonstration of the improvement to the science, the observations of M35 accumulated over about 13 years are
used to derive the proper motions. The results are compared with the ones using a conventional weighted fourth-order polynomial. In Section 5, some conclusions are drawn. 

\section{Instrument and observations} \label{sec:ins&obs}
To derive the geometric distortion (GD) solution for our astrometric programs, the open cluster M35 was observed for many years, and a large number of observations were accumulated with a 13 year time baseline. During the observation, a dithering strategy was arranged by a set of pointings with a step of about $\sim$1$'$. And the observations were obtained in I filter with exposure time of about 15--100 seconds. Details of the telescopes and CCD cameras are shown in Table~\ref{table:Instrumental Details}, and the details of the observations are shown in the first four columns of Table~\ref{table:Observation Details}. 

\begin{table*}
	\centering
	\small
	\caption{The specifications of the telescopes and the CCDs.}
	\begin{tabular}{ccccc}
		%\toprule[1pt]
		%\toprule[1pt]
		\hline
		\hline
		\makebox[0.13\textwidth][c]{\multirow{2}*{Parameter}} &\makebox[0.13\textwidth][c]{1.0-m (Kunming)}& \makebox[0.13\textwidth][c]{1.0-m (Kunming)}  &\makebox[0.13\textwidth][c]{0.8-m (Yaoan)}& \makebox[0.13\textwidth][c]{2.4-m (Lijiang)}\\ &\makebox[0.13\textwidth][c]{CCD\#1}& \makebox[0.13\textwidth][c]{CCD\#2}  &\makebox[0.13\textwidth][c]{CCD\#3}& \makebox[0.13\textwidth][c]{CCD\#4}
		\\
		\hline
Approximate focal length & 1330 cm &1330 cm & 800 cm & 1920 cm\\
Diameter of primary morror & 100 cm & 100 cm & 80 cm & 240 cm\\
		Size of CCD array            & 2048 $\times$ 2048      & 4096 $\times$ 4112   & 2048 $\times$ 2048      & 1900 $\times$ 1900                \\
		Size of pixel                & 13.5$\mu $ $\times $ 13.5$\mu $     &15$\mu $ $\times$ 15$\mu$  & 13.5$\mu $ $\times $ 13.5$\mu $     &15$\mu $ $\times$ 15$\mu$ \\
		Approximate scale factor     &
		0\farcs209/pixel       & 0\farcs234/pixel     &
		0\farcs346/pixel       & 0\farcs286/pixel         \\
		effective FOV                & $7\farcm1 \times7\farcm1$                    & $16\farcm0 \times 16\farcm0$   & $11\farcm8 \times11\farcm8$                    & $9\farcm0 \times 9\farcm0$          \\
		observed Nights & 10                                   & 3   & 2                                   & 1                               \\
%Site & Kunming & Kunming & Yaoan & Lijiang\\
		%\toprule[1pt]
		\hline
	\end{tabular}
	\label{table:Instrumental Details}
\end{table*}

\begin{table*}
\centering
\caption{Details of the observations for Section~\ref{sec:ins&obs} and the statistics of the results in terms of precision for Section~\ref{Comparison}. The first fifth columns are the specification of the observations, include CCD, observational epoch, the number and the exposure time of each observations set, the observational zenith distance~(ZD) and the seeing. The last six columns are the precisions derived from the conventional weighted fourth-order polynomial (subscript 1), the precisions derived from the PPT method (subscript 2), and the premium rates ($\mathcal{P}$) of precision by using PPT in two directions respectively. Note that the precisions are given in mas and they are derived from the stars brighter than 14 Gmag. More Details about the precision~($\sigma$) and the premium rate ($\mathcal{P}$) can be found in Section~\ref{Comparison2}.}

\begin{tabular}
{p{3em}p{4em}cccc*{6}{p{1.5em}<{\centering}}}
%{p{4em}p{3em}p{0.5em}cp{1.5em}p{1.5em}p{1.5em}p{1.5em}
%p{2em}p{2em}p{1.5em}p{1.5em}p{1.5em}p{1.5em}p{2em}p{2em}}
%{p{4em}p{3em}p{0.5em}cp{1.8em}p{1.8em}p{1.8em}p{1.8em}
%p{2em}p{2em}p{1.8em}p{1.8em}p{1.8em}p{1.8em}p{2em}p{2em}}
\hline
\hline
\multicolumn{1}{c}{CCD} &\multicolumn{1}{c}{Date} & \multicolumn{1}{c}{Exp-time} &  ZD & Seeing & \multicolumn{1}{c}{$\sigma_{\alpha 1}$} & \multicolumn{1}{c}{$\sigma_{\delta 1}$} & \multicolumn{1}{c}{$\sigma_{\alpha 2}$} & \multicolumn{1}{c}{$\sigma_{\delta 2}$}  &$\mathcal{P}\sigma{_\alpha}$ & $\mathcal{P}\sigma_{\delta}$ \\
  &   &     & ($^{\circ}$)   &  &&&& && \\
\hline
\multirow{10}*{CCD\#1}        &2009.01.07 &48$\times$100s       & \ 1.4 - 25.2  & $1.47\arcsec$-$1.57\arcsec$    &4.2  &4.3  &2.7  &3.3  & 36$\%$ & 23$\%$
\\
        &2010.11.30 &56$\times$80s       & \ 4.7 - 38.2   & $1.48\arcsec$-$1.57\arcsec$   &4.1  &4.4  &2.8  &2.6 & 32$\%$ & 41$\%$
\\
        &2011.02.24 &54$\times$60s       &\ 0.7 - 22.6  & $1.43\arcsec$-$1.54\arcsec$    &3.8  &4.6  &2.7  &2.6 &29$\%$ & 43$\%$
\\
        &2012.12.11 &49$\times$60s       &\ 0.7 - \, 7.9  & $1.49\arcsec$-$1.55\arcsec$      &3.5  &5.0  &2.6  &3.1 &26$\%$ & 38$\%$
\\
        &2015.02.12 &60$\times$80s       &\ 0.7 -  20.5  & $1.40\arcsec$-$1.49\arcsec$    &3.4  &3.8  &2.3  &2.0 &32$\%$ & 47$\%$
\\
        &2016.03.01 &49$\times$30s       &\ 0.6 -  11.7  & $1.45\arcsec$-$1.52\arcsec$    &4.2  &6.2  &3.0  &3.0 &29$\%$ & 52$\%$
\\
        &2017.11.12 &51$\times$40s       &\ 1.6 -  26.2  & $1.46\arcsec$-$1.57\arcsec$    &3.2  &3.6  &2.3  &2.2  &28$\%$ & 39$\%$
\\
        &2018.11.02 &79$\times$60s       & 12.0 -  43.1\, & $1.36\arcsec$-$1.57\arcsec$   &3.0  &3.6  &2.0  &1.8  &33$\%$ & 50$\%$
\\
        &2019.10.22 &67$\times$60s       &\ 0.7 -  16.3   & $1.39\arcsec$-$1.53\arcsec$    &4.5  &4.7  &3.2  &2.6 &29$\%$ & 45$\%$
\\
        &2019.11.23 &49$\times$60s       &\ 9.6 -  25.2   & $1.50\arcsec$-$1.56\arcsec$     &3.7  &4.8  &3.0  &2.9 &19$\%$ & 40$\%$
\\
\hline
\multirow{3}*{CCD\#2}        &2018.11.13 &52$\times$60s       &\ 0.6 -  16.5   & $1.64\arcsec$-$1.72\arcsec$    &3.2  &5.0  &0.9  &1.0 &72$\%$ & 80$\%$
\\
        &2021.01.15 &33$\times$30s       &\ 7.4 -  17.5   & $1.70\arcsec$-$1.93\arcsec$     &3.9  &6.2  &0.9  &0.9 &77$\%$ & 85$\%$
\\
        &2021.12.09 &96$\times$60s       &\ 0.6 -  24.1  & $1.73\arcsec$-$1.90\arcsec$     &4.5  &4.9  &1.6  &1.6 &64$\%$ & 67$\%$
\\
\hline
\multirow{2}*{CCD\#3}        &2019.10.12 &23$\times$60s       & 20.9 -  34.6\,  & $1.73\arcsec$-$1.97\arcsec$   &4.5  &5.3  &1.2  &1.2 &73$\%$ & 77$\%$
\\
        &2019.11.27 &40$\times$60s       &\ 2.6 -  14.2   & $1.89\arcsec$-$2.27\arcsec$     &4.6  &6.5  &1.7  &1.8 &63$\%$ & 72$\%$
\\
\hline
CCD\#4        &2013.02.06 &47$\times$15s       &\ 2.5 - \, 9.8  & $2.08\arcsec$-$2.15\arcsec$       &6.4  &8.5  &4.1  &4.6 &36$\%$ & 46$\%$
\\
\hline
  average&             &      &   &   &  &  &  & &42$\%$ & 53$\%$
\\
\hline
\end{tabular}
\label{table:Observation Details}
\end{table*}

\section{Method}
\label{section:Method}

\subsection{Preliminary Reduction}
Firstly, the standard procedures including de-bias and flat fielding are performed. The pixel positions of stars are achieved by a two-dimensional Gaussian centering. Secondly, we match the star's pixel positions with \textit{Gaia} DR3~\citep{Gaia2021A&A} and calculate their topocentric apparent positions at the observational epoch considering all the astrometric effects and the atmosphere refraction effect. Note that the Gaia sources, of which the renormalised unit weight errors~(RUWEs) are larger than 2.0 have been excluded. Thirdly, through the central projection the standard coordinate of each matched star is computed and we calibrate the charge transfer efficiency~(CTE) effect and the differential colour refraction~(DCR) effect of the observations according to Lin et al.~(\citeyear{Lin2020MNRAS}). Fourthly, an average GD solution is derived by the method proposed by Peng et al.(\citeyear{Peng2012AJ}) and applied to the stars' pixel positions. Finally, the positional results via a weighted fourth-order polynomial can be derived, which can absorb the dynamic distortion effect at the observational epoch.  And the positional precision~($\sigma_p$) of each star can be estimated through a sigmoidal function
describing the relation between its Gmag and its positional
precision~(similar as the fitted curve of the first column of Figure.1 in~\citealt{Lin2019MNRAS}). For each observation set, its precision premium curve (PPC) is derived~~(see the first column of Figure.3 in~\citealt{Lin2019MNRAS}), and the relative positional precision ($\sigma_r$) as a function of the angular seperation between any two stars can be estimated.

\subsection{Methodology}
After the preliminary reduction, the DCR effects, the CTE effects, and the average GD effects of the observations have been corrected. And the major errors left on the stars' pixel positions result from three sources: 1) stochastic wander of the star source caused by the atmospheric turbulence; 2) dynamic distortion caused by the variation of thermal-induced or flexure-induced in telescope's optics during the observation; 3) certering noise based on a source's SNR. And the PPT method, which corrects the positional bias caused by turbulence, called turbulence errors as follows, and also dynamic distortion, is interpreted as follows. 

Since Gaia DR3 provides accurate positions without atmospheric turbulence errors, for a dense sky region, such as an open cluster, the accurate positions can be used to form a net to reduce turbulence errors. Specifically, if the region of the net is small enough, the turbulence errors of the pixel positions in the net are similar, and therefore the turbulence errors in the center of a net can be significantly reduced by relating the pixel positions (affected by turbulence) of the stars in the net to their topocentric apparent positions (unaffected by turbulence) in the process of the determination for the plate model. Each position of a star on the observation is taken as the central position of the net and its observed apparent position would be derived via a weighted first-order or second-order polynomial, which depends on the number of the high-SNR reference stars. Note that the polynomial can also absorb the dynamic GD effect effectively since it varies little in a small region. By considering the contribution of positional precision ($\sigma_p$) and the relative positional precision ($\sigma_r$) as a function of the angular seperation between the reference star and the central position of the net, the weight of each reference star is set to $1/(\sigma_p^{2}-\sigma_{p0}^{2}+\sigma_r^{2}/2)$, where $\sigma_{p0}$ is the relative positional precision of the highest-SNR reference star on each observations.

\begin{figure*}
\includegraphics[width=0.9\textwidth]{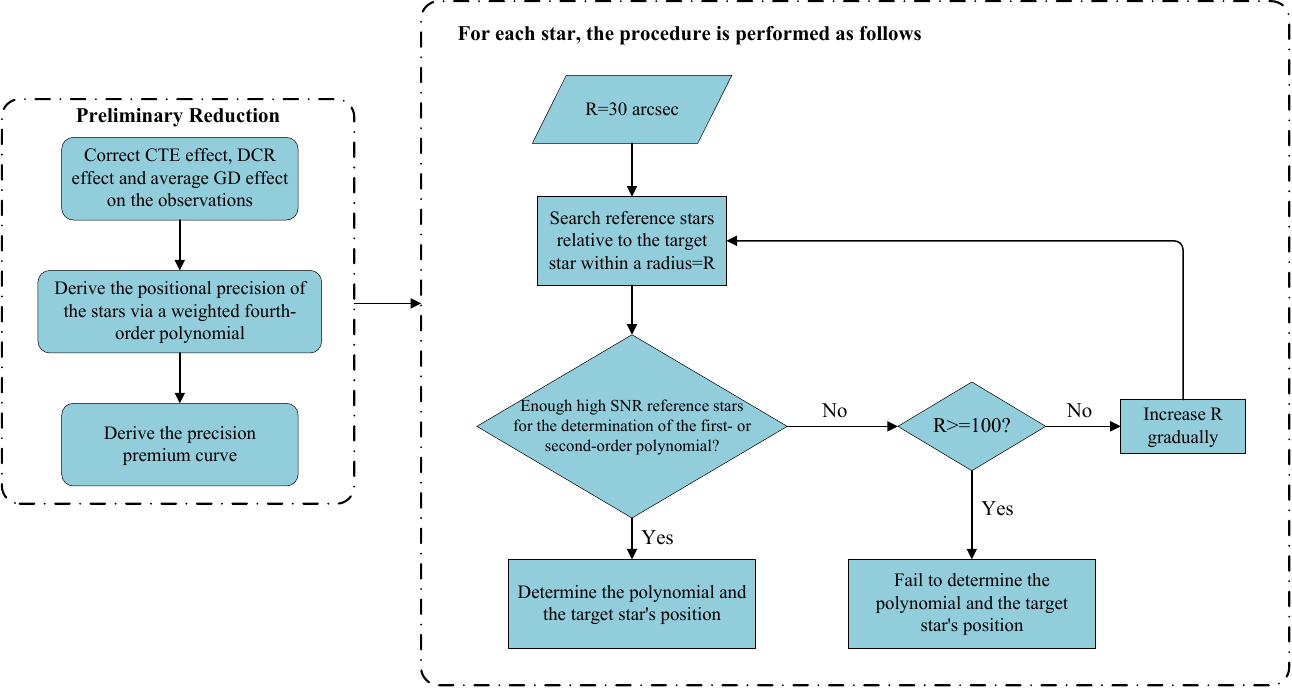}
\caption{Diagram of the PPT procedure.}
\label{Fig1}
\end{figure*}

Initially, the region is set to 30 arcsec when the reduction of each star is started. If there are not enough reference stars for the determination of the plate model, the region will be increased gradually until the condition is met. And the procedure of PPT is visualized in Figure~\ref{Fig1}. 

\section{Comparison}\label{Comparison}

In this section, the results derived via the PPT method are compared with, respectively, the results derived via the traditional premium precision (PP) application and the results derived via a conventional weighted fourth-order polynomial~\citep{Lin2019MNRAS} in terms of precision. Moreover, we pay attention to how many targets can be measured via the PPT method, i.e. its applicability, for the observation of an open cluster. For the consideration of the numbers of the reference stars in FOV and the observational zenith distance, two observation-sets are chosen to be processed in this section, which are taken on 2018.11.13 and 2021.01.15 respectively with CCD\#2. 

\subsection{Comparison with the traditional PP application}\label{Comparison2}
The traditional PP application takes only single star as reference, which is required to have a high SNR image on the observation, while the PPT method measures a target 
relative to several reference stars, including the high-SNR images and the low-SNR images as well. The results derived via the two methods are compared in terms of precision as follows.   

\begin{figure*}
\centering
\subfloat[2018.11.13]{\includegraphics[width=0.48\textwidth]{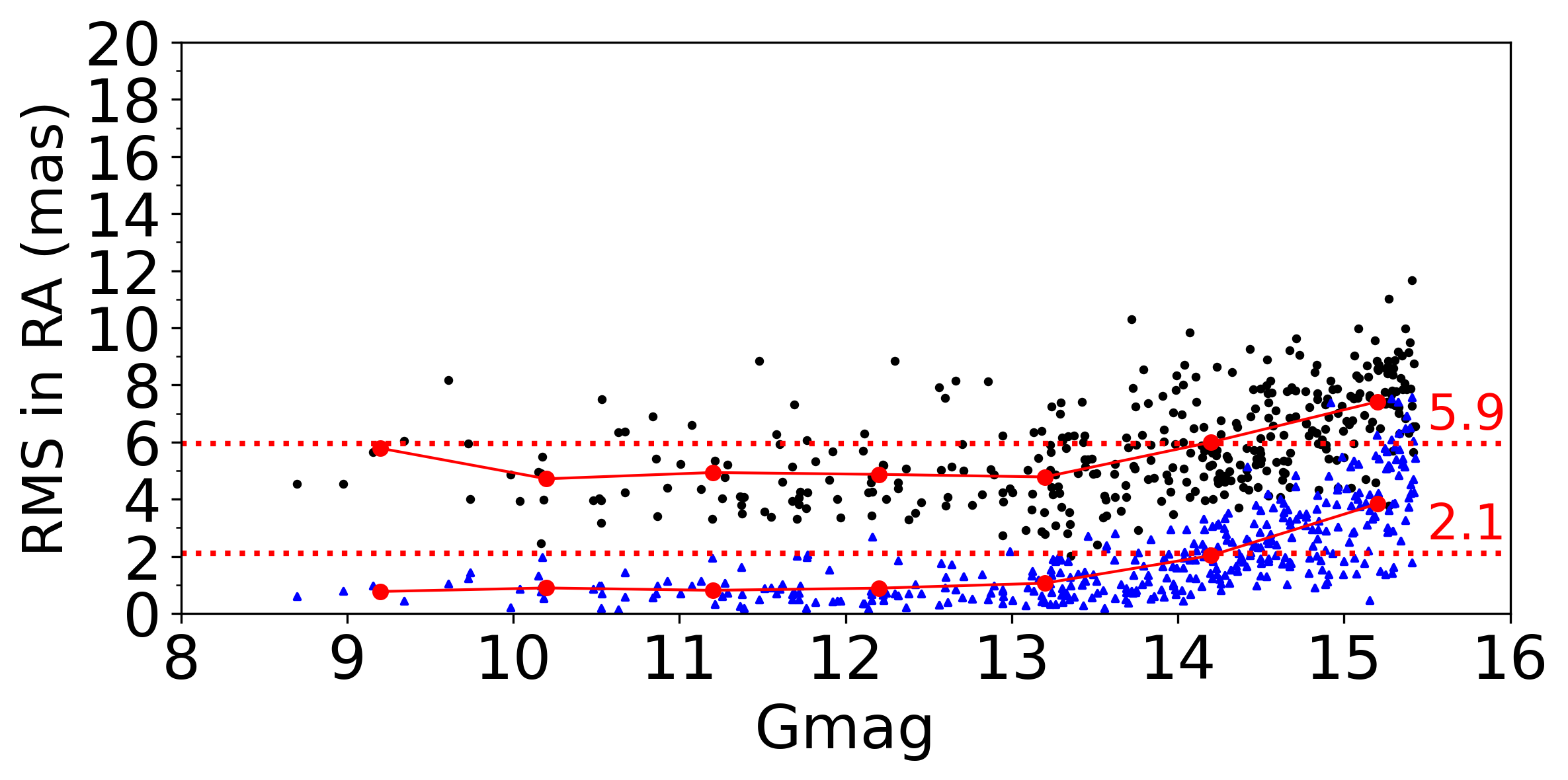} \includegraphics[width=0.48\textwidth]{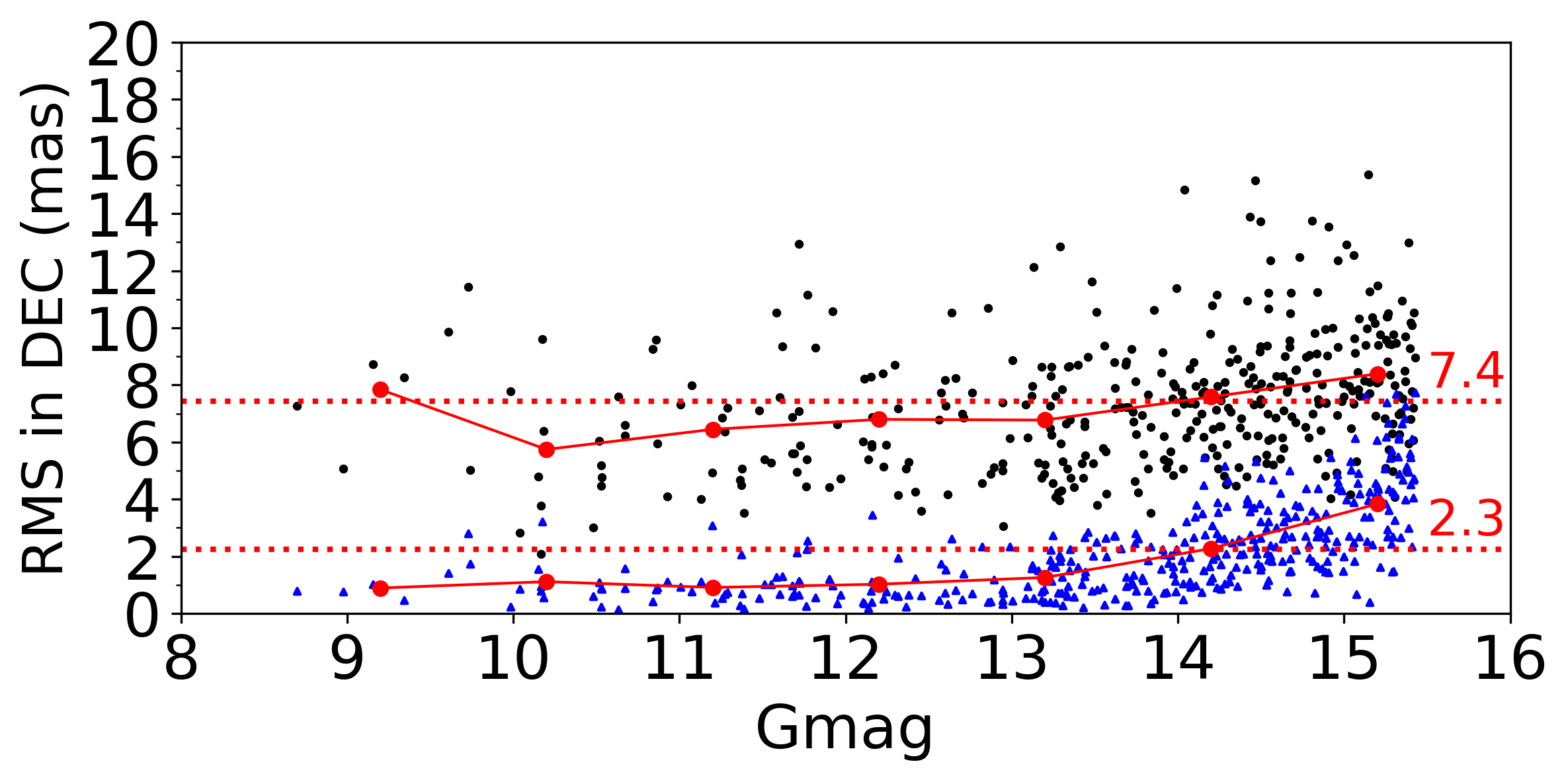}}\\
\subfloat[2020.01.15]{\includegraphics[width=0.48\textwidth]{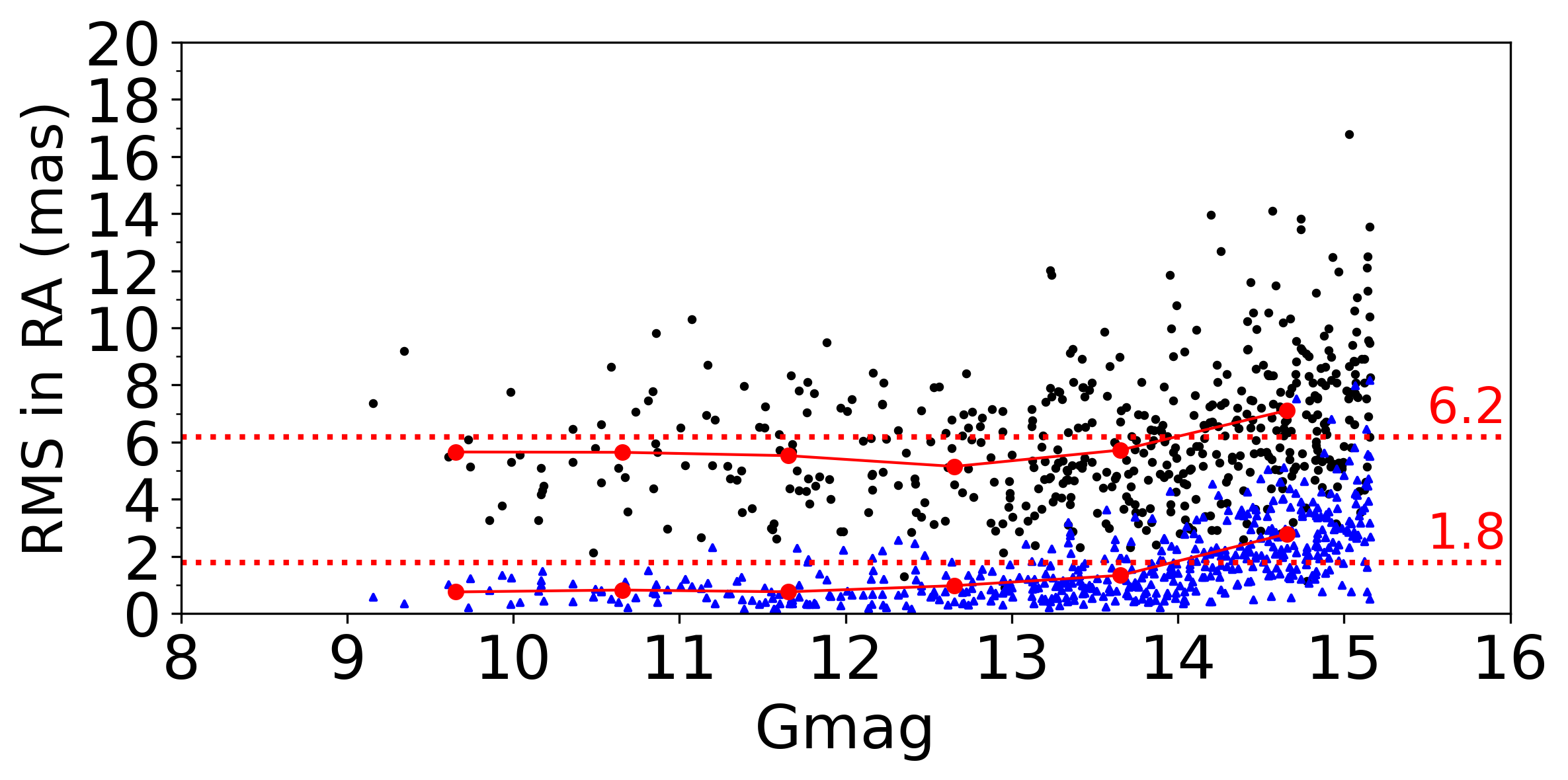} \includegraphics[width=0.48\textwidth]{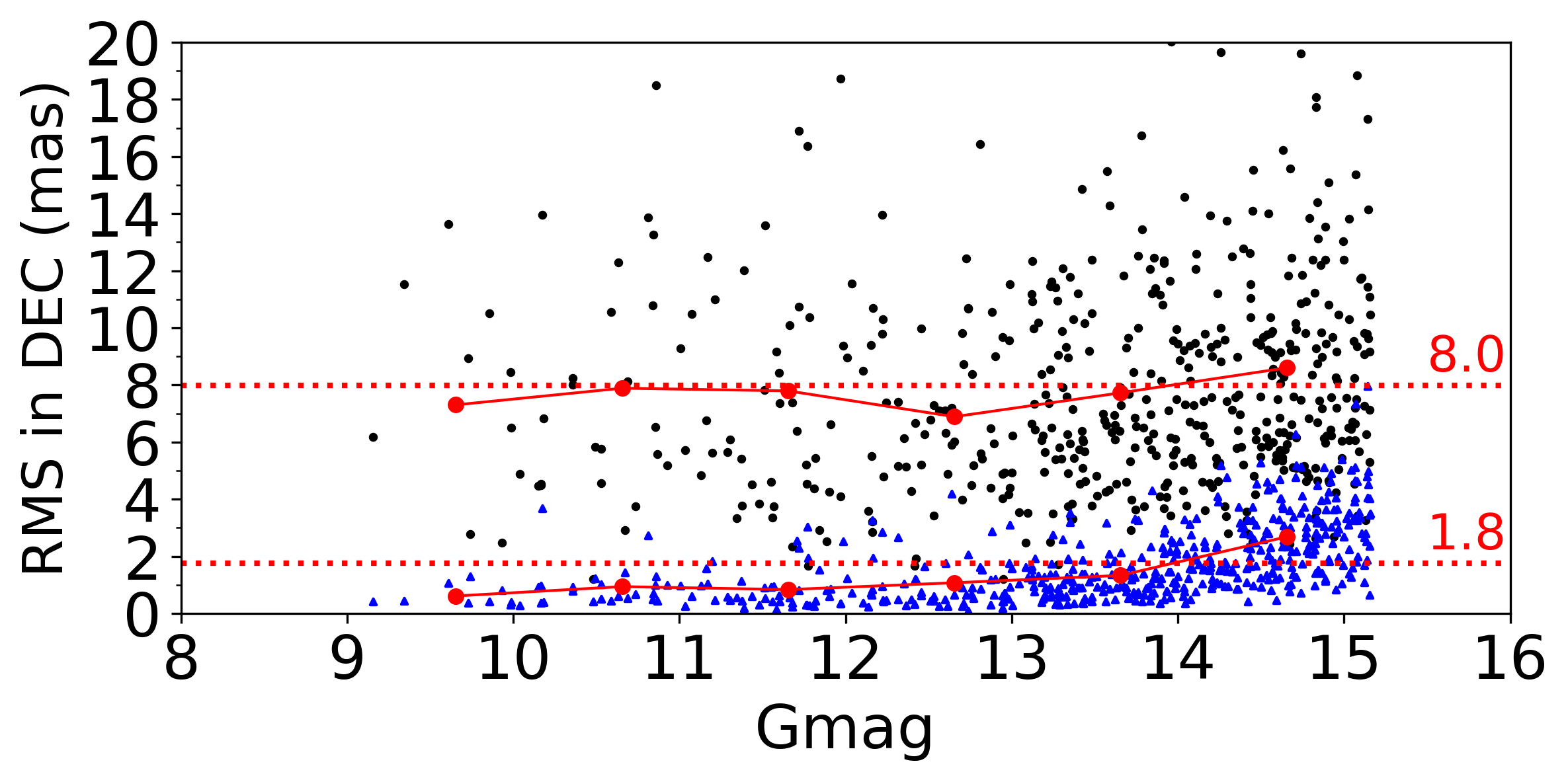}}\\
\caption{The results of the precisions derived via the traditional PP application (black circle) and the PPT method (blue triangle) for 2018.11.13 and 2020.01.15 respectively. Average of each bin of 1 Gmag is computed and marked as a red circle. The red dotted line marks the average precision for all the results of each method.}
\label{Fig2}
\end{figure*}

To ensure the traditional PP application to achieve high-precision results, both the target star and the reference star should have positional precisions~($\sigma_p$) which are not worse than 10 mas. Moreover, the target star would be measured relative to the nearest reference star within 60-arcsec radius. For the same target stars, the results of the two observation-sets derived via the two methods are shown in Figure~\ref{Fig2}. The PPT method exhibits a significant improvement over the tradition PP application for high-SNR stars in terms of precision. Moreover, the deviation of the precisions between RA and DEC directions becomes smaller by using the PPT method since it takes more stars as reference for the determination of the plate model.

\begin{figure*}
  \centering
  \subfloat[2018.11.13]{\label{figur:3.1}\includegraphics[width=0.48\textwidth]{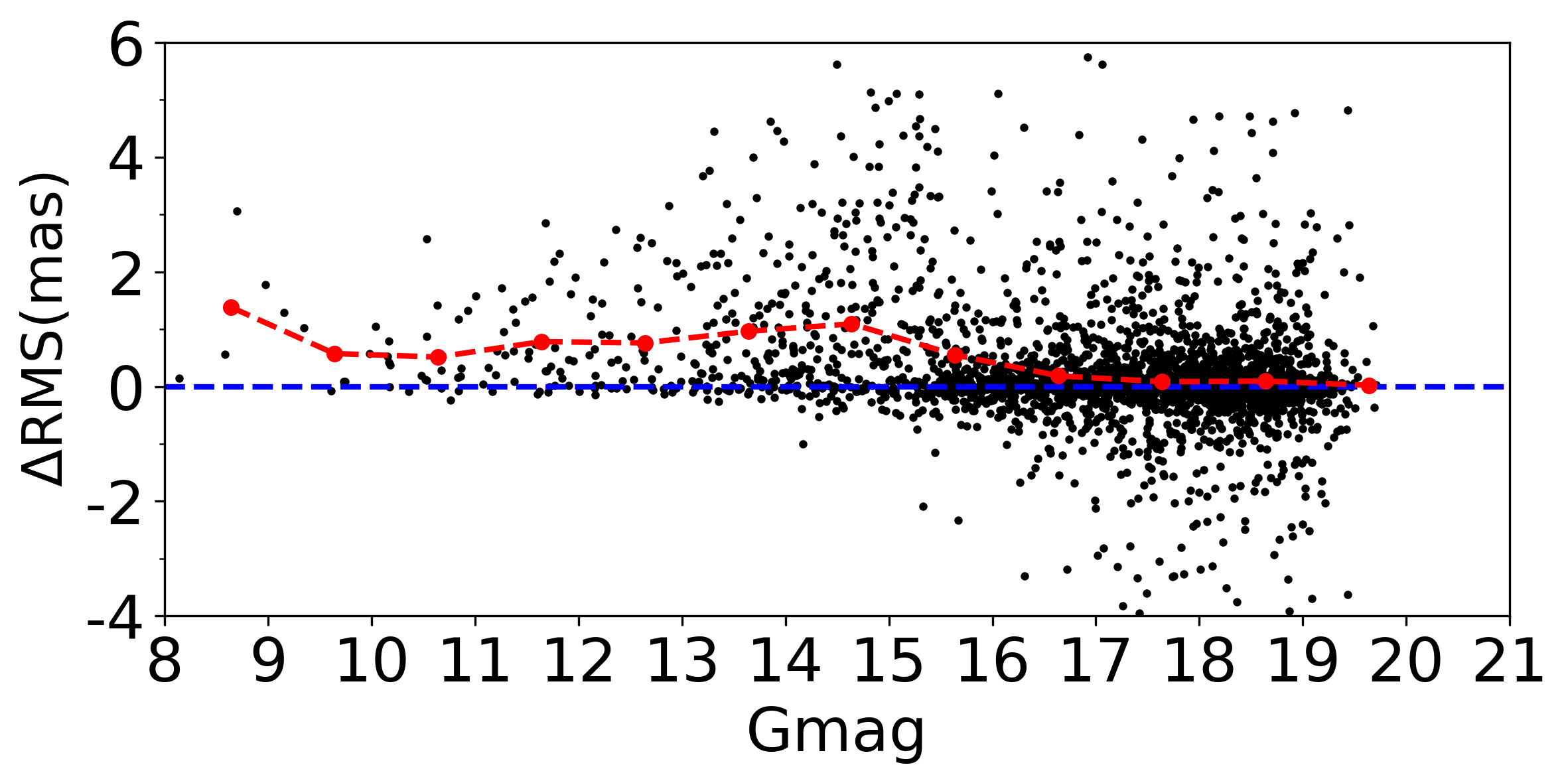}}
  \subfloat[2020.01.15]{\label{figur:3.2}\includegraphics[width=0.48\textwidth]{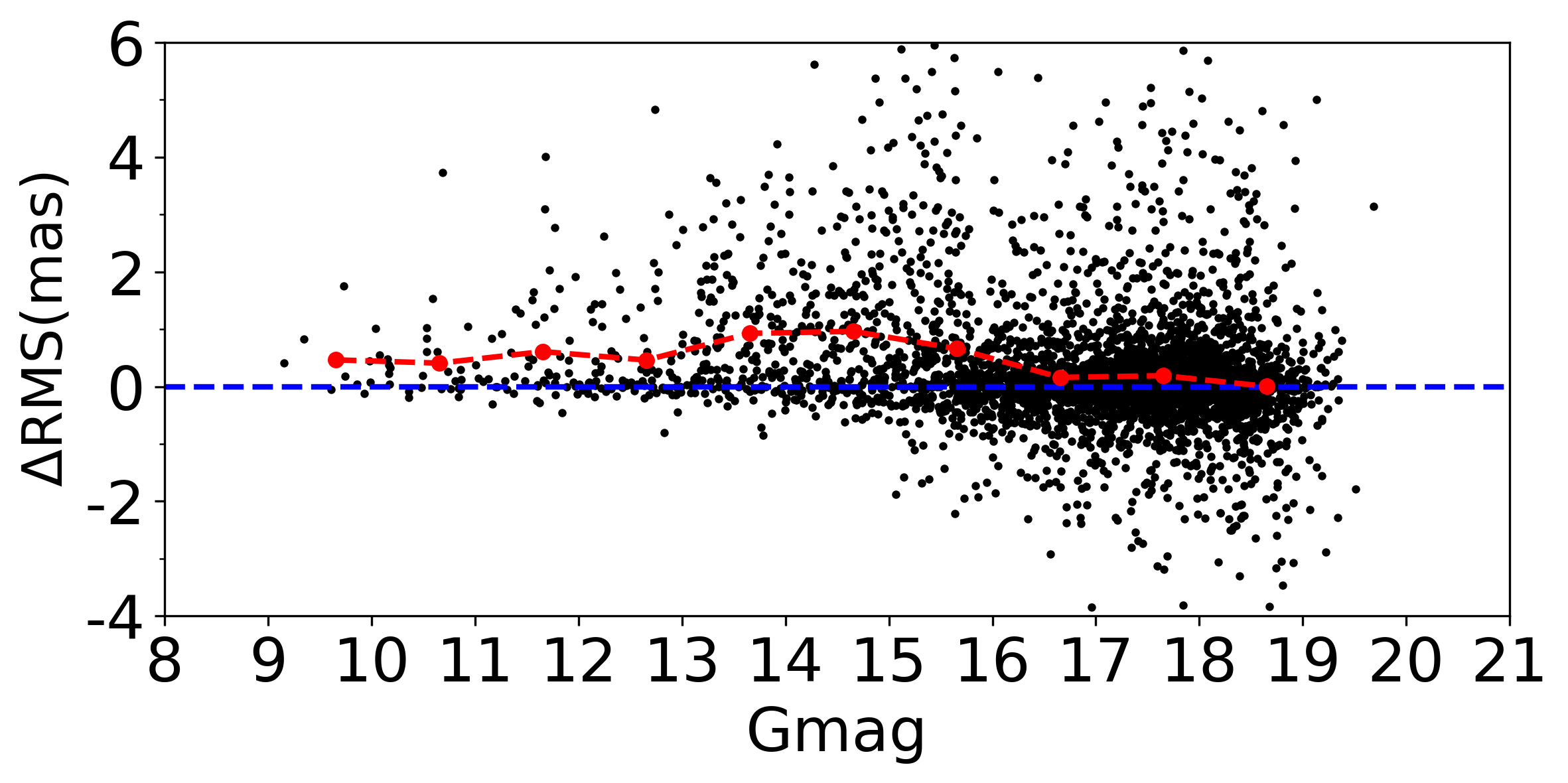}}\\
  \caption{The differences of the precisions ($\sqrt{\sigma_ {\rm{\alpha cos\delta}}^{\rm{2}}+\sigma_{\rm{\delta}}^{\rm{2}}}$) by using only high-SNR reference stars and by using not only high-SNR reference stars but also low-SNR reference stars when measuring the same stars via the PPT method. Average of each bin of 1 Gmag is computed and marked as a red circle.}
\label{Fig3}
\end{figure*}

To further look into how the low-SNR reference stars affect the results of the PPT method, the PPT method is performed again by using only the high-SNR reference stars, the precision of which are not worse than 30 mas. The derived results are compared with the results by using the high-SNR reference stars and low-SNR reference stars as well. The differences of the precisions are shown in Figure~\ref{Fig3}. For the high-SNR stars, improvement at a certain extent can be seen when considering all the stars within the effective radius of the PPT method. 

%\begin{figure*}
%  \centering
%  \subfloat[]{\label{figur:1}\includegraphics[width=0.48\textwidth]{result2_12SD_RA_5.9_2.1.png}}
%  \subfloat[]{\label{figur:2}\includegraphics[width=0.48\textwidth]{result2_12SD_DEC_7.4_2.3.png}}\\
%  \subfloat[]{\label{figur:1}\includegraphics[width=0.48\textwidth]{result4_12SD_RA_6.2_1.8.png}}
%  \subfloat[]{\label{figur:2}\includegraphics[width=0.48\textwidth]{result4_12SD_DEC_8.0_1.8.png}} 
%  \caption{The results of the precisions derived via the traditional PP method (black circle) and the PPT method (blue triangle) for 2018.11.13 and 2020.01.15 respectively. Average of each bin of 1 Gmag is computed and marked as a red circle. The red dashed line marks the average precision for all the results of each method.}
%\label{Fig2}
%\end{figure*}

\subsection{Comparison with a conventional weighted fourth-order polynomial}\label{Comparison2}

\begin{figure*}
\subfloat[2018.11.13]{\label{figur:3.1}\includegraphics[width=0.96\textwidth]{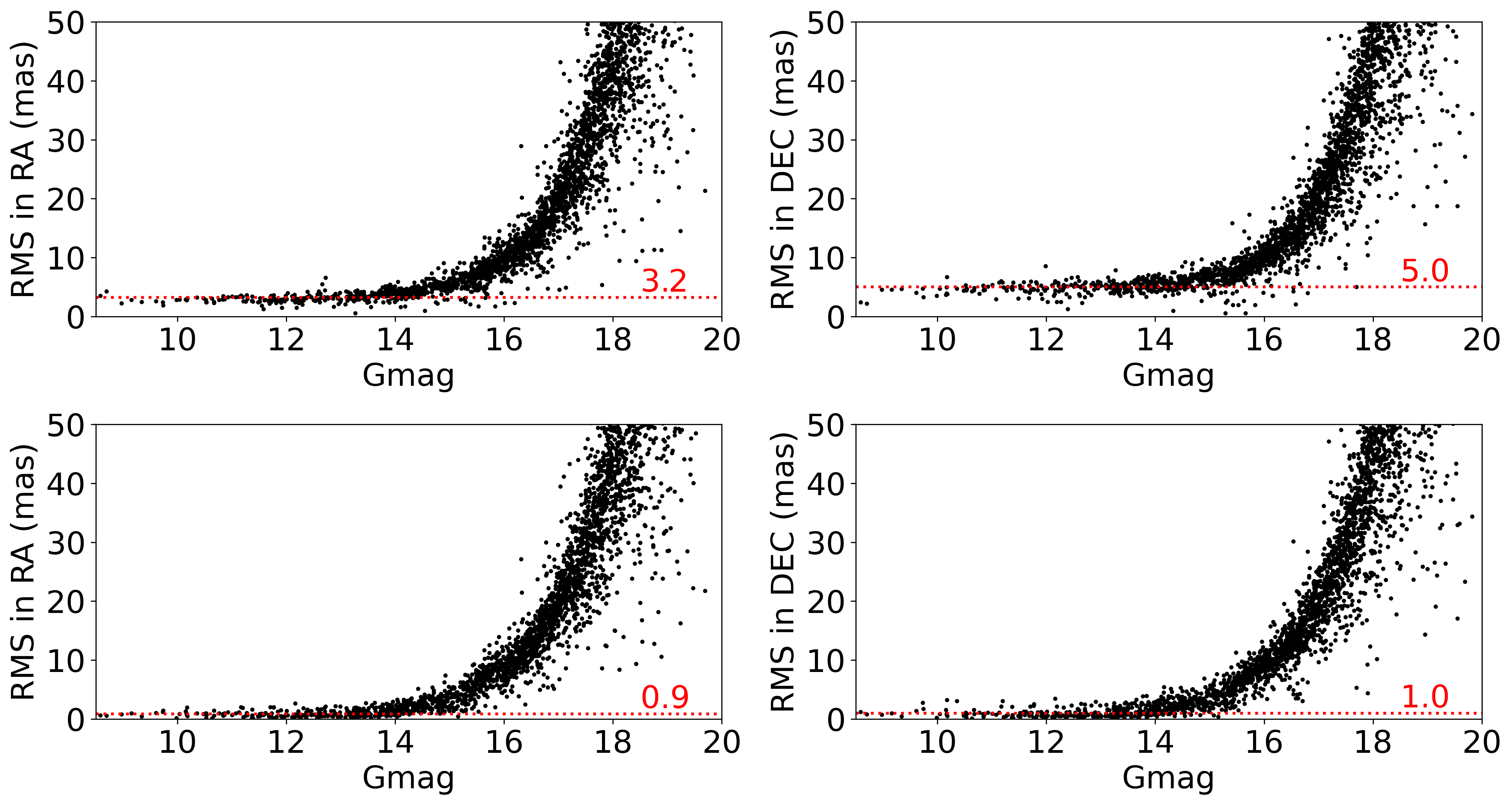}}\\
\subfloat[2020.01.15]{\label{figur:3.2}\includegraphics[width=0.96\textwidth]{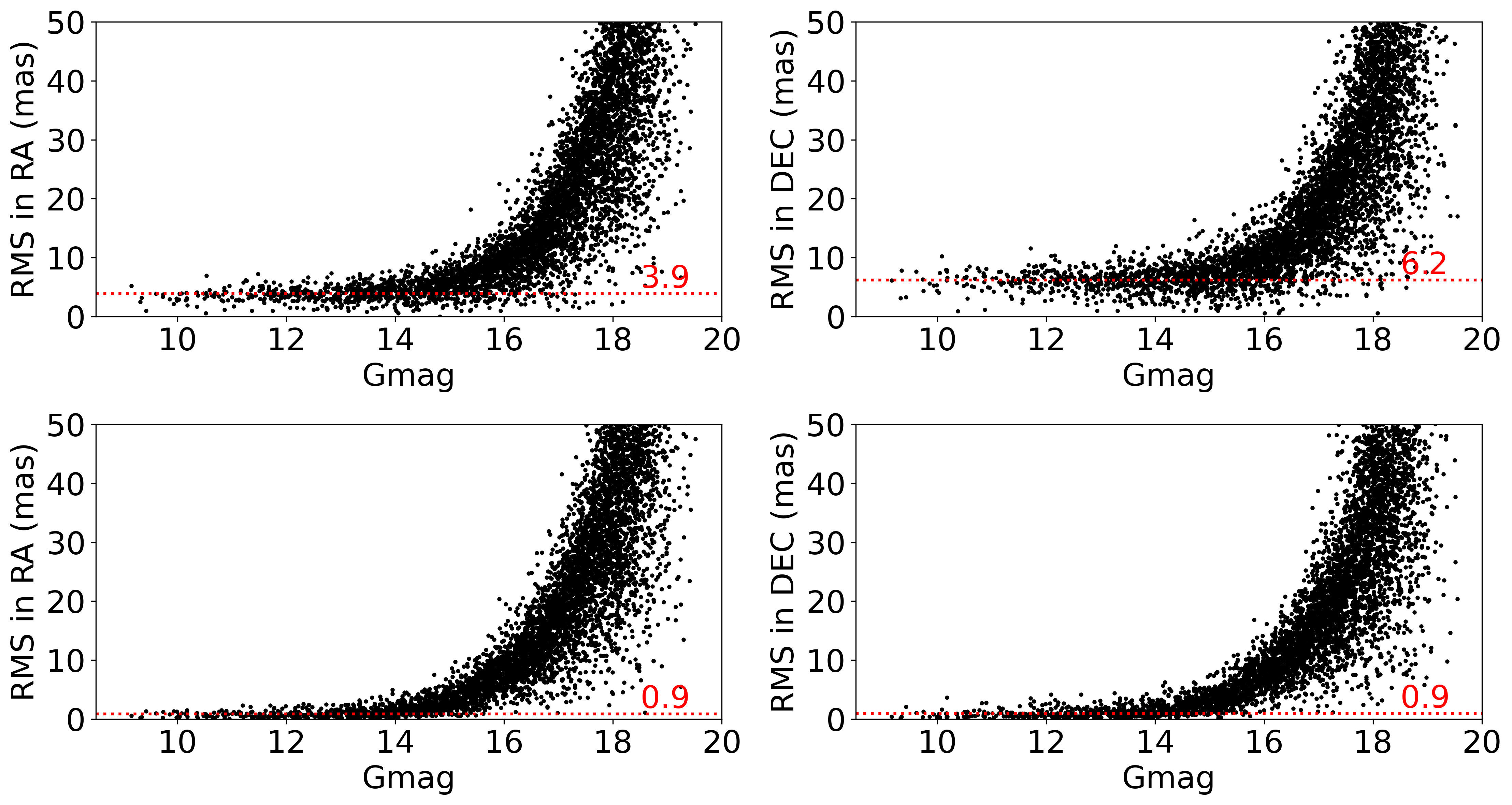}}
\caption{The results of the precisions derived via a conventional weighted fourth-order polynomial and the PPT method. For each panel, the results derived via a conventional method are shown in the first row and the results derived via the PPT method are shown in the second row. The red dotted line marks the average precision for the stars brighter than 14 Gmag.}
\label{Fig4}
\end{figure*}

A conventional weighted fourth-order polynomial~\citep{Lin2019MNRAS} are performed for the same observation-set. The detailed results of the precision via a weighted fourth-order polynomial and the PPT method are shown in Figure~\ref{Fig4}. The statistics of all the results are shown at the final six columns of Table~\ref{table:Instrumental Details}.

PPT exhibits a significant improvement over the high-order polynomial for bright stars in terms of precision, which uses all the reference stars in the FOV for its determination. To further demonstrate the relative improvement level, we define the premium rate $\mathcal{P}$ in percentage: 
\begin{equation}\label{equ1}
\mathcal{P}=(R_{conv}-R_{new})/R_{conv}\times 100\%,
\end{equation}
%$$
%\begin{array}{c}
%\mathcal{P}=(Q_{conv}-Q_{new})/Q_{conv},
%\end{array}
%$$
where $R_{conv}$ is the result derived via the conventional method and $R_{new}$ is the result derived via the new method. 

\begin{figure}
\centering
\includegraphics[width=0.8\columnwidth]{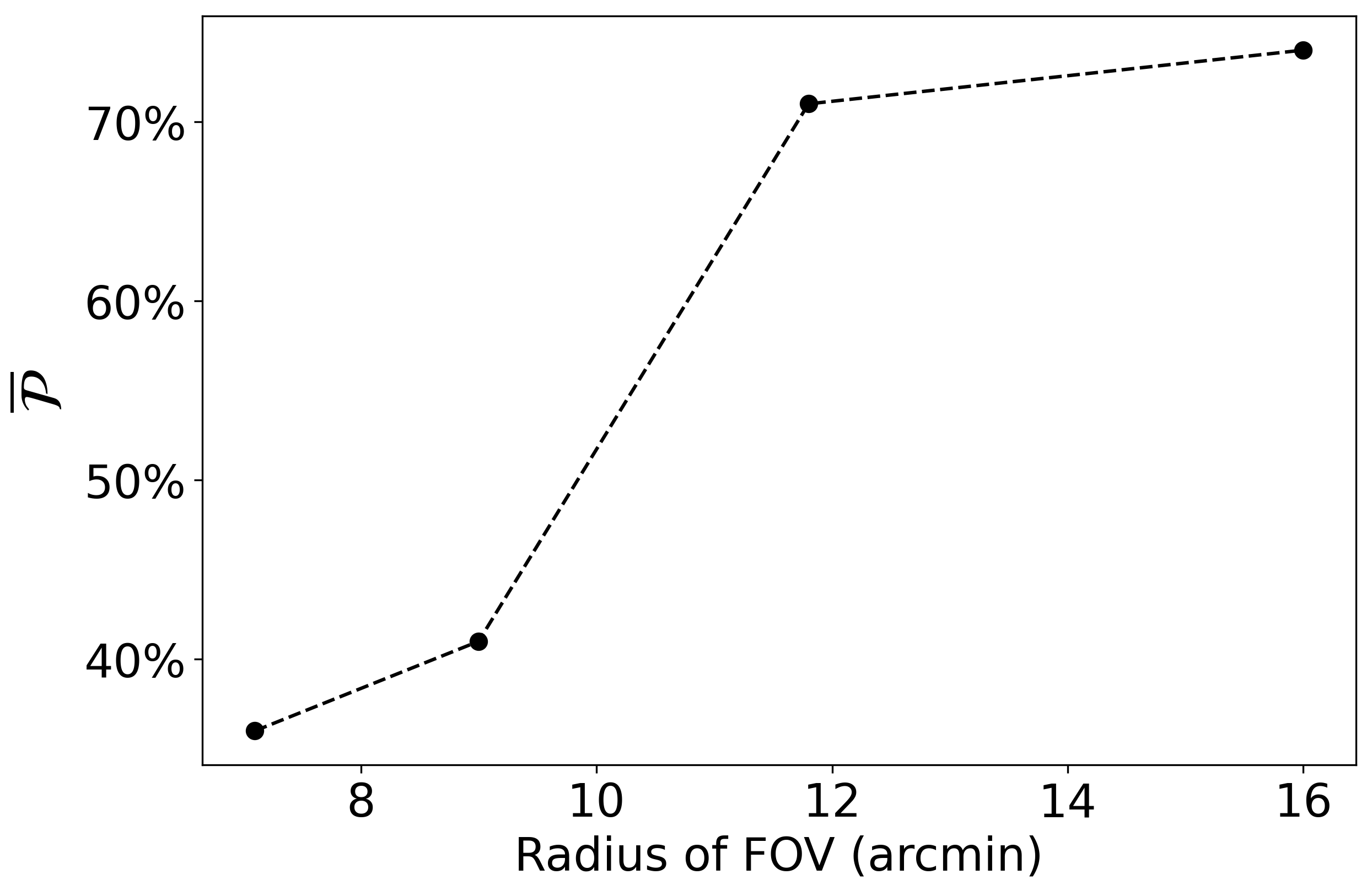}
\caption{the average premium rate of the precision ($\overline{\mathcal{P}}$) as a function of the side of FOV}
\label{Fig8}
\end{figure}

Instead of using a conventional weighted high-order polynomial, the premium rates $\mathcal{P}$ of precision by using the PPT method are shown in the last two columns of Table~\ref{table:Instrumental Details}. The average of the premium rate $\mathcal{P}$ of precision is 42\% in RA direction and 53\% in DEC direction, respectively. We also calculate the average premium rate ($\overline{\mathcal{P}}$) of precision for each CCD and $\overline{\mathcal{P}}$ as a function of the side of FOV is shown in Figure~\ref{Fig8}. It is remarkably, that the improvements by using CCD\#2 and CCD\#3 are more significant than the other two CCDs. We think the greater improvement should be attributed to their larger FOV which can capture more reference stars to perform the PPT method. We also explore the relationship between the positional precision~($\sigma$) and, respectively, $T^{-0.5}$ and the seeing in the Table~\ref{table:Observation Details}, where $T$ is the exposure time of the observation. But no obvious trends are found.

Another improvement the PPT method bestows should be
that, the deviation of the precision and the accuracy between RA and DEC directions becomes smaller. The major contribution for the improvement should be that, the PPT method has corrected the turbulence errors on the observation effectively, which has a great impact on the centering results at X or Y direction or both. And we think there may be another contribution: the PPT method also corrects most of the instrumental and propagation effects, which are common for star images sufficiently close together, since the PPT method performs local measurement in a small region. 

%since CCD\#2 has the largest FOV and the lowest readout noise, which is cooled by liquid nitrogen.
%And the contribution should be that the PPT method performs local measurement in a small region, in which 
%most of the instrumental and propagation effects are common for star images sufficiently close together.

\begin{figure*}
  \centering
  \subfloat[2018.11.13]{\label{figur:1}\includegraphics[width=0.48\textwidth]{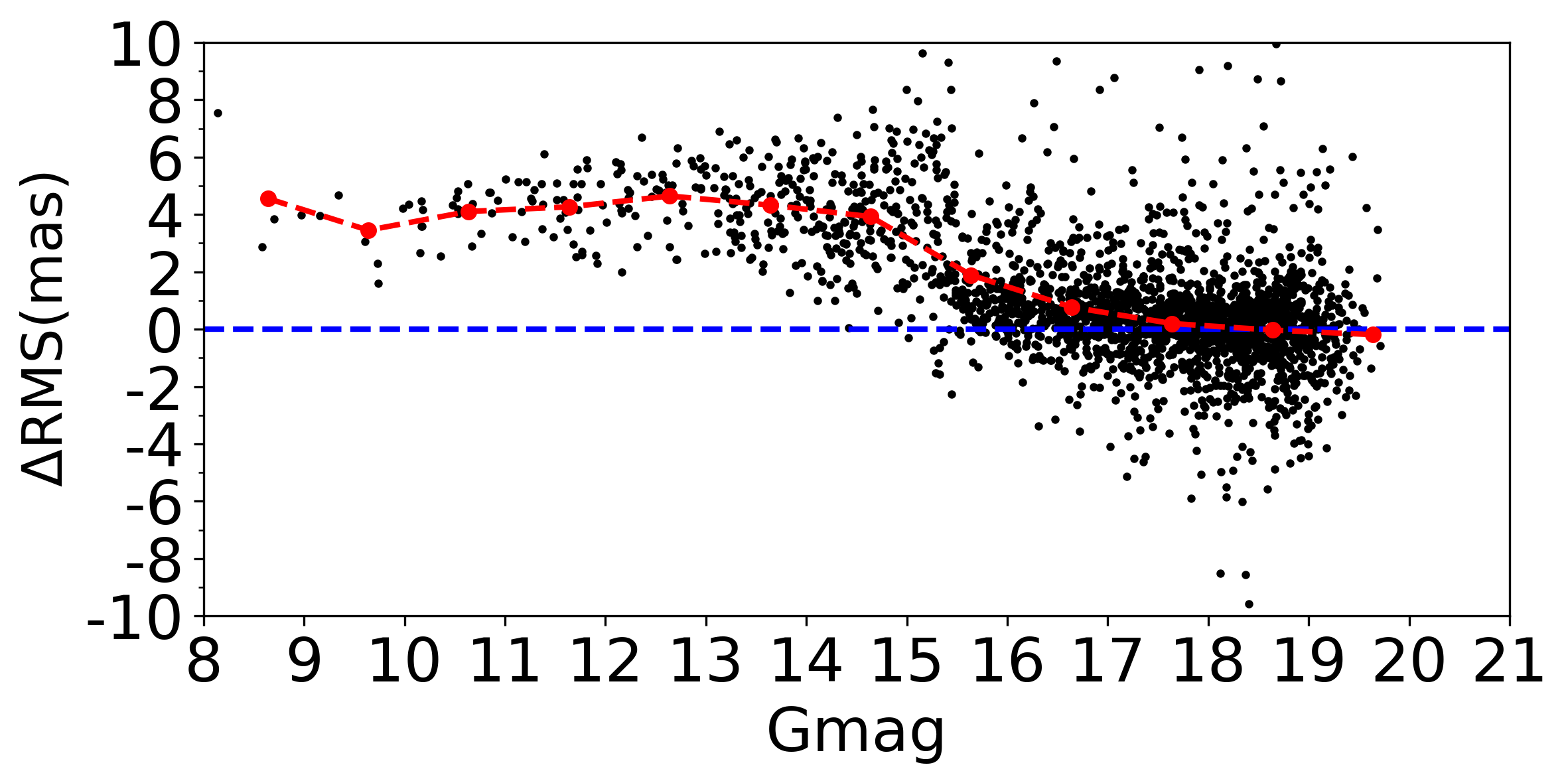}}
  \subfloat[2020.01.15]{\label{figur:2}\includegraphics[width=0.48\textwidth]{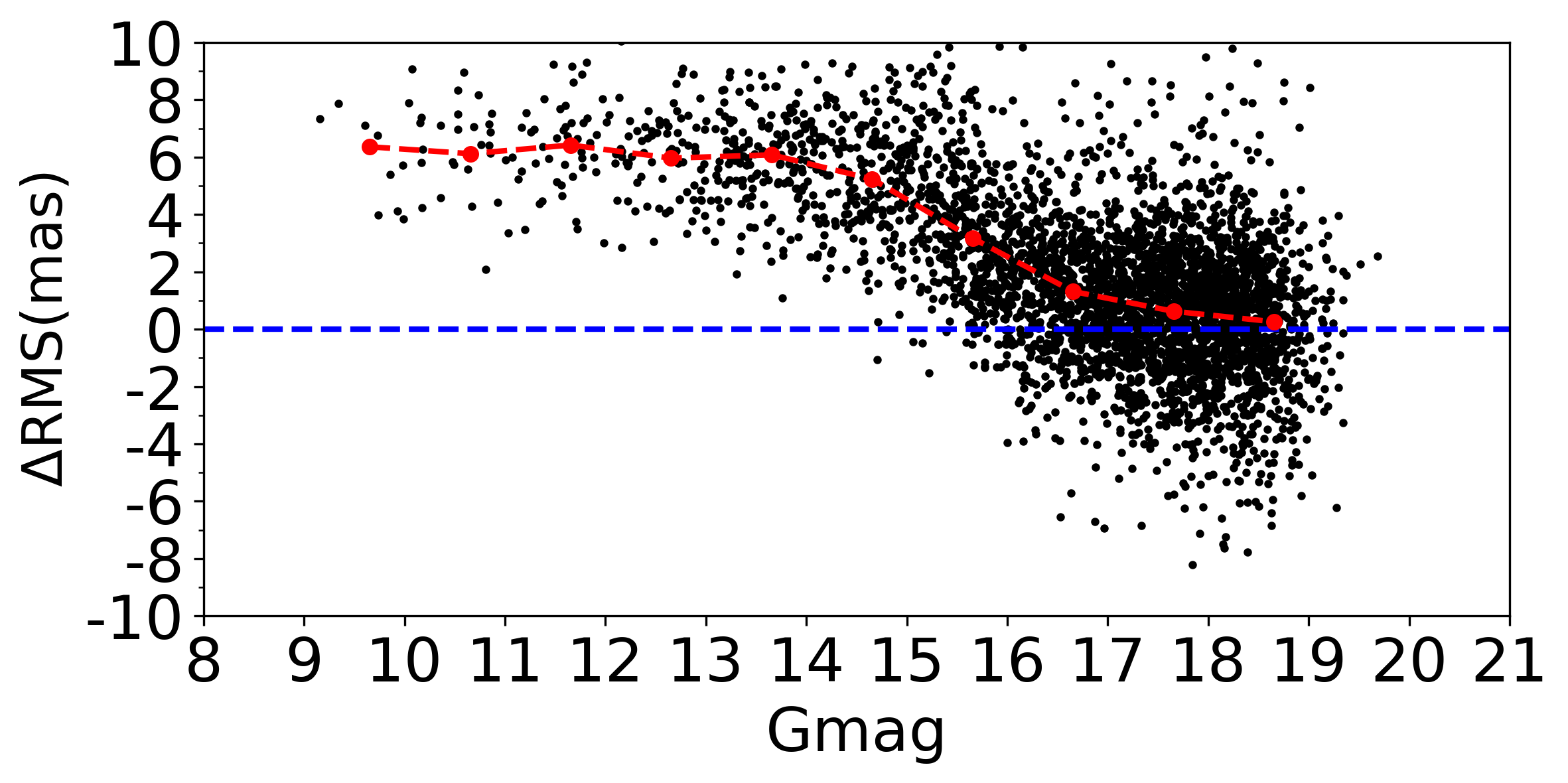}}\\
  \caption{The differences of the precisions ($\sqrt{\sigma_ {\rm{\alpha cos\delta}}^{\rm{2}}+\sigma_{\rm{\delta}}^{\rm{2}}}$) for the observations captured with 4k CCD at Kunming on 2018.11.13~(Figure~\ref{Fig4}, panel b) and 2021.1.15~(Figure~\ref{Fig4}, panel b), via two different methods. Average of each bin of 1 Gmag is computed and is marked as a red circle.}
\label{Fig5}
\end{figure*}

To further look into the performance of the PPT method, we make a comparison between the precisions derived by the two methods. The difference of the precisions is shown in Figure~\ref{Fig5}. To our surprise, even for stars as faint as about 17 Gmag, the PPT method has the potential to improve their average precisions. However, the improvement level seems to decrease for stars fainter than 14 Gmag, of which the centering noise might become the dominant contribution to positional measurement. 

Finally, to investigate the applicability of the PPT, we compare how many targets can be measured via the PPT method and a conventional weighted fourth-order polynomial. As shown in Table~\ref{table:applicability}, there is about 3\% stars failed to be measured when the effective radius of the PPT method is restricted to be within 100 arcsec. However, if the effective radius is slightly increased to 110 arcsec, all the stars can be measured via the PPT method. We also check the new results and find that the precision of the high-SNR stars are slightly improved compared with a weighted fourth-order polynomial.  

\begin{table}
\centering
\small
\caption{The number of the derived precisions via different methods.}
\begin{tabular}{ccc}
%\toprule[1pt]
%\toprule[1pt]
\hline
\hline
\makebox[0.23\columnwidth][c]{Method} &\makebox[0.15\columnwidth][c]{2018.11.13} &\makebox[0.15\columnwidth][c]{2021.01.15}\\ 
\hline
PPT & 3206 & 5771  \\
fourth-order & 3334 & 5901 \\
\hline
\end{tabular}
\label{table:applicability}
\end{table}

\section{Application}
As a demonstration of the improvement to the science, the observations of M35 accumulated over about 13 years are used to derive the proper motion~(PM). The procedure is interpreted as follows. 

Firstly, for each star, its astrometric position can be calculated by using SOFA library~\citep{Wallace1998}. Its observed astrometric position can be derived by adding the residual, i.e. the ($O-C$)~(the observed minus the computed), which is derived in Section 4. Then for each observational epoch, the parallax effects are corrected for the observed astrometric positions. For a star with N observations, we have the quadruplet ($\alpha_N$,$\delta_N$,$t_N$,$\sigma_N$), where ($\alpha_N$,$\delta_N$) is its observed astrometric positions, $t_N$ is its observational epoch and $\sigma_N$ is the positional uncertainties. To measure
the PM of this star in RA and DEC direction, we used a 
weighted least-squares to fit a straight line to the data points
($\alpha_N$,$t_N$,$\sigma_N$) and ($\delta_N$,$t_N$,$\sigma_N$) respectively. We progressively improved the fit by rejecting outliers or badly measured observations. Note that, we only measure the PM of the star which has at least 100 data points, with at least 5 years of time baseline. These conditions must be satisfied at every stage of the fitting
and rejection process.

\begin{figure*}
    \begin{minipage}{\textwidth}
    \centering
	\includegraphics[width=0.75\columnwidth]{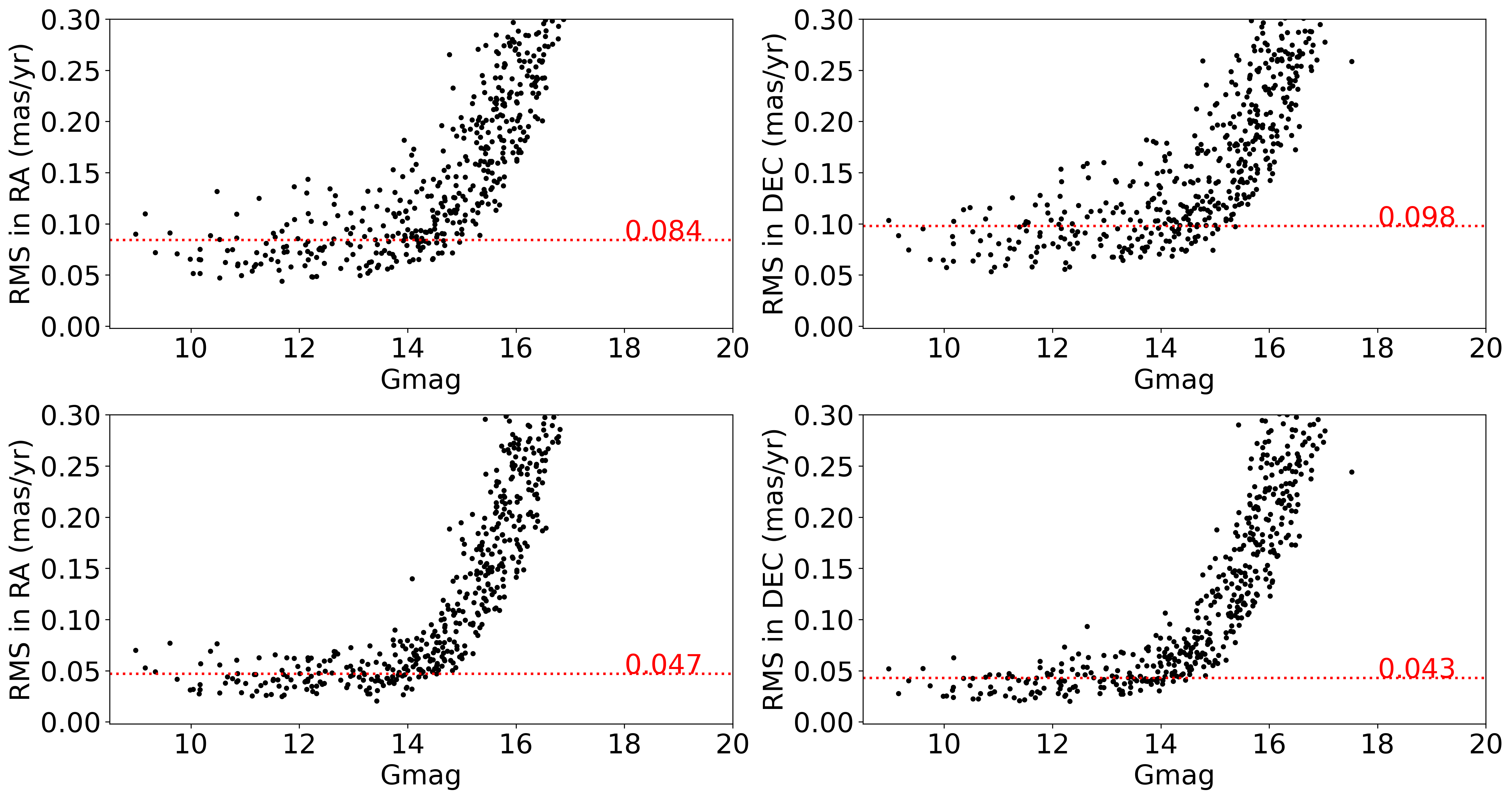}
    \end{minipage}
    \caption{The precisions derived via a conventional weighted fourth-order polynomial and the PPT method. The results derived via a conventional method are shown in the first row and the results derived via the PPT method are shown in the second row. The red dotted line marks the average precision for the stars brighter than 14 Gmag.}
    \label{Fig6}
\end{figure*}

The precisions of the derived PMs via the PPT method and the conventional weighted fourth-order polynomial are shown in Figure~\ref{Fig6}. Similar conclusion can be drawn as Section 4. Compared with the conventional method, the PPT method exhibits a significant improvement especially in DEC direction. What is more important, the precisions of the two directions by the PPT method stay almost at the same level.

\section{Conclusion} \label{sec:conclusion}
Based on the precision premium curve (PPC),
we proposed a high-precision astrometric solution called precision premium transformation (PPT) in this paper, which
takes advantage of high similarity of the positional bias caused mostly by turbulence (called turbulence error in this paper) in a small region and the high-precision Gaia reference stars in the region to reduce 
the turbulence errors on the observations, through a weighted solution. The PPT method is applied to the observations of an open cluster, M35, which have been a calibration field for our astrometric program for many years. Through systematic analysis, the PPT method exhibits significant advantages in terms of not only precision but also applicability when a target is in an area of high stellar density. Instead of using the conventional weighted fourth-order polynomial, the average of the relative improvement level of precision in percentage, or the called premium rate $\mathcal{P}$ defined in this paper, is 42\% in RA direction and 53\% in DEC direction, respectively, when using the PPT method. And it is found that $\mathcal{P}$ increases with larger FOV of CCD. The opportunity to apply the PPT method would improve with the advent of more deep and dense astrometric catalogues and telescopes with a large FOV in the future. 

Another improvement the PPT method bestows should be that, the deviation of the precision between RA and DEC direction becomes smaller. We think the contribution should be that, the PPT method performs local measurement in a small region, in which not only turbulence errors but also most of the instrumental and propagation effects are common for star images
sufficiently close together. Since there is a definite trend to install mosaic instruments with increasing numbers of chip elements~\citep{Zheng2022A&A} for current or future optical imaging telescopes, intensive researches are needed to derive average correction models of the complex systematic effects. After applying average correction models, many of the residual systematic effects that affect positional precision can further be absorbed via the PPT method.

\section*{Acknowledgements}

This work was supported by the National Key R\&D Program of China (Grant No. 2022YFE0116800), by the China Manned Space Project (Grant No. CMS-CSST-2021-B08), by the National Natural Science Foundation of China (Grant Nos. 11873026, 11273014, 12203019), by the Natural Science Foundation of Guangdong Province (Grant No. 2023A1515011270), by the Joint Research Fund in Astronomy (Grant No. U1431227) and by the Scientific Research Starting Foundation of Guangdong Ocean University. This work has made use of data from the European Space Agency (ESA) mission Gaia (\url {https://www. cosmos.esa.int/gaia}), processed by the Gaia Data Processing and Analysis Consortium (DPAC; \url{hppts://www.cosmos.esa.int /web/gaia/dpac/consortium}). Funding for the DPAC has been provided by national institutions, in particular the institutions participating in the Gaia Multilateral Agreement.

%%%%%%%%%%%%%%%%%%%%%%%%%%%%%%%%%%%%%%%%%%%%%%%%%%


\bibliographystyle{aa}
\bibliography{mypaper} % if your bibtex file is called example.bib

%\begin{thebibliography}{99}
%\end{thebibliography}

%%%%%%%%%%%%%%%%%%%%%%%%%%%%%%%%%%%%%%%%%%%%%%%%%%

%%%%%%%%%%%%%%%%% APPENDICES %%%%%%%%%%%%%%%%%%%%%

%\appendix

%%%%%%%%%%%%%%%%%%%%%%%%%%%%%%%%%%%%%%%%%%%%%%%%%%

\end{document}